\documentclass[12pt]{article}

\usepackage{epsf}
\usepackage{amsmath}
\usepackage{graphics}
\usepackage{amsfonts}
\usepackage{amssymb}
\usepackage{latexsym}
\usepackage{color}
\usepackage[dvips]{graphicx}
\input{colordvi.tex}

\renewcommand{\thefootnote}{\#\arabic{footnote}}

\begin{document}
\setcounter{footnote}{0}

\begin{titlepage}

\begin{center}


\vskip .5in

{\Large \bf
Extension of local-type inequality for
the higher order correlation functions}

\vskip .45in

{\large
Teruaki Suyama$^1$
and 
Shuichiro Yokoyama$^2$
}

\vskip .45in

{\em
$^1$
  Research Center for the Early Universe, Graduate School
  of Science, The University of Tokyo, Tokyo 113-0033, Japan
 \vspace{0.2cm} \\
$^2$Department of Physics and Astrophysics, Nagoya University,
Aichi 464-8602, Japan}

\end{center}

\vskip .4in

\begin{abstract}
For the local-type primordial perturbation, it is known that there is 
an inequality between the bispectrum and the trispectrum.
By using the diagrammatic method, we develop a general formalism to systematically 
construct the similar inequalities up to any order correlation function.
As an application, we explicitly derive all the inequalities up to six and eight-point
functions.
\end{abstract}
\end{titlepage}

\renewcommand{\thepage}{\arabic{page}}
\setcounter{page}{1}
\renewcommand{\thefootnote}{\#\arabic{footnote}}

\section{Introduction}
Primordial non-Gaussianity has been attracting attention as a powerful probe to discriminate
many existing inflation models (for example, see \cite{Komatsu:2009kd,Bartolo:2010qu}).
Statistics of zero-mean Gaussian fluctuations can be characterized by
a variance which corresponds to two-point function.
Hence, the non-Gaussianity of fluctuations
can be linked to the higher-order correlation functions. 
A lot of studies have been done for the bispectrum, i.e. three-point function of the
primordial perturbations.
For the so-called local type perturbation, the bispectrum is completely characterized
by a single non-linearity parameter $f_{\rm NL}$~\cite{Komatsu:2001rj}.
The current bound on $f_{\rm NL}$ is $ -10 < f_{\rm NL} < 74$ at $95\%$ confidence level~\cite{Komatsu:2010fb}.

In the last few years, trispectrum is becoming an important observable as well.
For the local type non-Gaussianity, the trispectrum is specified by two non-linearity parameters
$\tau_{\rm NL}$ and $g_{\rm NL}$~\cite{Boubekeur:2005fj,Byrnes:2006vq}.
The current bounds on $\tau_{\rm NL}$ and $g_{\rm NL}$ are respectively $-0.6 <  \tau_{\rm NL}/10^4 \le 3.3$
and $-7.4 < g_{\rm NL} / 10^5 < 8.2 $ at $95\%$ confidence level~\cite{Smidt:2010ra}.
An interesting fact that boosts the importance of the study of the trispectrum is 
that $\tau_{\rm NL}$ has a minimum determined by $f_{\rm NL}$ \cite{Suyama:2007bg},
\begin{equation}
\tau_{\rm NL} \ge \frac{36}{25}f_{\rm NL}^2. \label{ft}
\end{equation}
It was shown in Ref.~\cite{Kogo:2006kh} that Planck satellite can measure $\tau_{\rm NL}$ down to $560$.
Therefore, future detection of $f_{\rm NL}\simeq 30$ which is the central value of the current bound
implies that we should also detect non-vanishing $\tau_{\rm NL}$.
Detection of both $f_{\rm NL}$ and $\tau_{\rm NL}$ (and possibly $g_{\rm NL}$ as well) enables us
to constrain or even pin down the inflation model and the origin of the primordial fluctuations.
In fact, in Ref.~\cite{Smidt:2010ra} the authors have discussed an observational constraint
on the ratio between $f_{\rm NL}$ and $\tau_{\rm NL}$
which is defined as $A_{\rm NL} \equiv \tau_{\rm NL} / (6f_{\rm NL}/5)^2$.

There exist a few studies that extend the inequality (\ref{ft}).
The authors of Refs.~\cite{Suyama:2010uj, Sugiyama:2011jt} considered one-loop corrections to (\ref{ft})
and found the one-loop effect appears as the scale dependent modification of the coefficient 
in front of $f_{\rm NL}^2$.
In Ref.~\cite{Yokoyama:2008by}, we have discussed that how many parameters one needs in order to
characterize higher order correlation functions of primordial curvature fluctuations and 
also introduced the non-linearity parameters for the five-point function. 
The non-linearity parameters up to the six-point function were introduced in \cite{Lin:2010ua,Meyers:2011mm}
and two inequalities between the six- and four-point function were derived \cite{Lin:2010ua}.

A potential importance and usefulness of the higher order correlation functions motivate
us to look for the similar inequalities like (\ref{ft}) for the higher order correlation
functions.
In this paper, we provide a general formalism to construct the inequalities  
among the non-linearity parameters for the higher order correlation functions.
To this end, we will adopt the diagrammatic approach developed in Refs.~\cite{Yokoyama:2008by,Byrnes:2007tm}
that turns out to be a convenient way to study the higher order correlation functions.
By using this method, in principle, we can systematically derive all the inequalities up 
to any-point function.

This paper is organized as follows.
In the next section, we briefly review the higher order correlation
functions of primordial curvature fluctuations, based on $\delta N$ formalism and the diagrammatic approach.
In section~\ref{sec:inequality},
we define the non-linearity parameters to characterize the higher order functions
and construct the series of inequalities by applying the Cauchy-Schwarz inequality.
 We also present all inequalities up to the six- and eight-point functions.
Section~\ref{sec:con} is spent on the conclusion of this paper.

\section{Higher order correlation functions of $\delta N$}

According to the $\delta N$ formalism \cite{Starobinsky:1986fx,Sasaki:1995aw,Sasaki:1998ug,Lyth:2004gb}, 
the curvature perturbation on the uniform energy density hypersurface
on super-horizon scales at $t=t_f$ is equal to the perturbation of the e-folding number evaluated from 
the flat hypersurface at $t=t_*$ to the uniform energy density hypersurface at $t_f$ at the same point:
\begin{equation}
\zeta (t_f,{\vec x})=\delta N (t_f,t_*;{\vec x}). \label{deltaN1}
\end{equation}
The initial time $t_*$ can be chosen arbitrary.
If we chose $t_*$ as the time slightly after the scale we are interested in leaves the Hubble horizon
during inflation, then $\delta N$ in Eq.~(\ref{deltaN1}) is sourced by the scalar field fluctuations.
Thus, $\zeta$ can be Taylor-expanded in terms of the scalar field fluctuations as
\begin{equation}
\zeta (t_f,{\vec x})=\sum \frac{1}{n!} N_{a_1 a_2 \cdots a_n} \delta \phi^{a_1}(t_*,{\vec x}) \cdots \delta \phi^{a_n}(t_*,{\vec x}),
\end{equation}
where $a_1,~a_2,\cdots$ run from $1$ to $p$ ($p$ is a number of light scalar fields that
acquire super-horizon scale fluctuations during inflation).
In the following, we assume that $\delta \phi^a$ are Gaussian variables.


The two-point function of $\zeta$ can be written as
\begin{equation}
\langle \zeta_{\vec k_1} \zeta_{\vec k_2} \rangle ={(2\pi)}^3 \delta ({\vec k_1}+{\vec k_2}) P_\zeta(k_1). 
\end{equation}
To leading order in the field fluctuations, $P_\zeta (k)$ is given by
\begin{equation}
P_\zeta (k)=N_a N_a P (k),~~~~~P(k)=\frac{2\pi^2}{k^3} {\left( \frac{H_*}{2\pi} \right)}^2,
\end{equation}
where summation over $a$ is assumed.

The three-point function of $\zeta$ can be written as
\begin{equation}
\langle \zeta_{\vec k_1} \zeta_{\vec k_2} \zeta_{\vec k_3} \rangle ={(2\pi)}^3 \delta ({\vec k_1}+{\vec k_2}+{\vec k_3}) B_\zeta(k_1,k_2,k_3). 
\end{equation}
To leading order in the field fluctuations, $B_\zeta$ is given by
\begin{equation}
B_\zeta (k_1,k_2,k_3)=\frac{N_a N_b N_{ab}}{{(N_c N_c)}^2} \left( P_\zeta (k_1) P_\zeta (k_2)+2~{\rm perms.} \right).
\end{equation}
Following literatures, the constant coefficient in front of the square of the power spectrum is 
written as \cite{Lyth:2005fi}
\begin{equation}
f_{\rm NL}=\frac{5}{6} \frac{N_a N_b N_{ab}}{{(N_c N_c)}^2}.
\end{equation}

The four-point function of $\zeta$ can be written as
\begin{equation}
\langle \zeta_{\vec k_1} \zeta_{\vec k_2} \zeta_{\vec k_3} \zeta_{\vec k_4} \rangle ={(2\pi)}^3 \delta ({\vec k_1}+{\vec k_2}+{\vec k_3}+{\vec k_4}) T_\zeta({\vec k_1},{\vec k_2},{\vec k_3},{\vec k_4}). 
\end{equation}
To leading order in the field fluctuations, $T_\zeta$ is given by
\begin{eqnarray}
T_\zeta({\vec k_1},{\vec k_2},{\vec k_3},{\vec k_4})&=&\frac{N_a N_{ab} N_{bc}N_c}{{(N_d N_d)}^3} \left( P_\zeta (k_1) P_\zeta (k_{12})P_\zeta(k_4)+11~{\rm perms.} \right) \nonumber \\
&&+\frac{N_a N_b N_c N_{abc}}{{(N_d N_d)}^3} \left( P_\zeta (k_1) P_\zeta (k_2) P_\zeta (k_3)+3~{\rm perms.} \right), 
\end{eqnarray}
where $k_{ij} \equiv | {\vec k_i}+{\vec k_j} |$.
Note that there appear two distinct terms that exhibit different wavenumber dependence.
As a consequence, we need two parameters to specify the tri-spectrum.
Following literatures, the two constant parameters are defined by \cite{Byrnes:2006vq}
\begin{equation}
\tau_{\rm NL}=\frac{N_a N_{ab} N_{bc}N_c}{{(N_d N_d)}^3},~~~~~g_{\rm NL}=\frac{25}{54} \frac{N_a N_b N_c N_{abc}}{{(N_d N_d)}^3}.
\end{equation}

\begin{figure}[tbp]
  \begin{center}
    \includegraphics{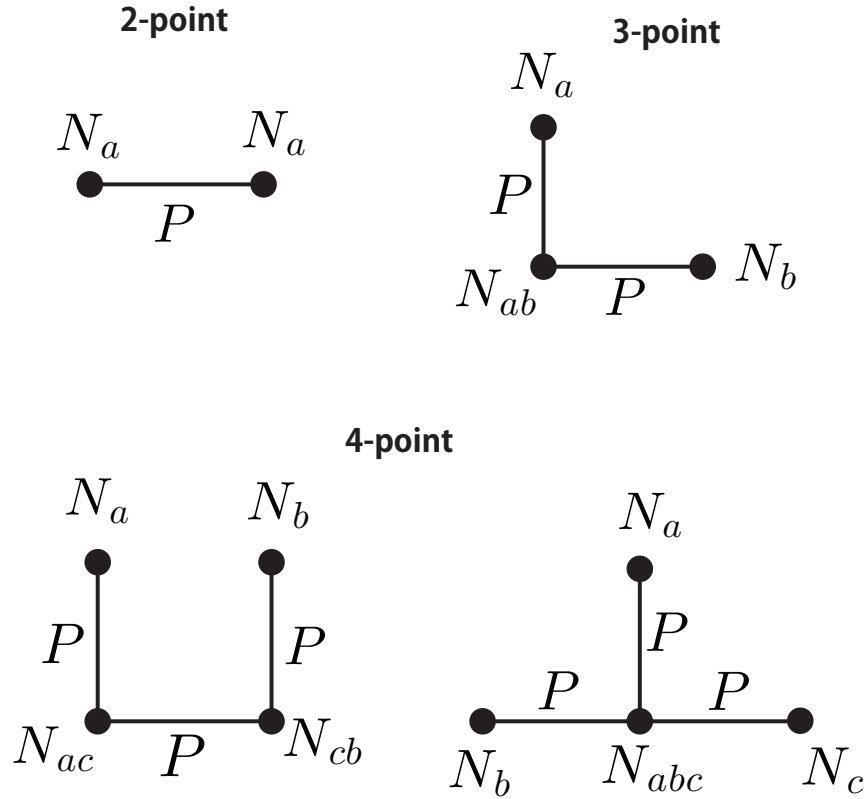}
  \end{center}
  \caption{The tree diagrams corresponding to the power spectrum, bispectrum and trispectrum, based on $\delta N$ formalism.
  We find that we need two parameters in order to characterize the trispectrum.}
  \label{fig:diagram_2.eps}
\end{figure}

We can further proceed to higher order correlation functions as follows.
The leading order of the $n$-point function is given by a sum of several 
distinct terms which are products of $(n-1)$ power spectra and exhibit different wavenumber dependence.
According to the diagrammatic method~\cite{Yokoyama:2008by}, 
each of these leading terms has a corresponding connected tree
diagram that consists of $n$ vertices and $(n-1)$ lines connecting two vertices.
Tree diagrams for the two-,~three- and four-point functions are shown in Fig.~\ref{fig:diagram_2.eps}.
%
%
%
In reverse, given a tree diagram with $n$ vertices, we can reconstruct the corresponding term that constitutes
the $n$-point function as follows.
First, we assign a different wavenumber $\{ {\vec k_1},\cdots,{\vec k_n} \}$ to each vertex of the diagram,
where $\{ {\vec k_1},\cdots,{\vec k_n} \}$ are the arguments of the $n$-point function with the
constraint ${\vec k_1}+\cdots+{\vec k_n}=0$.
Next, we assign a wavenumber to each line in the diagram, too.
Removing a line from the diagram yields two respectively connected sub-diagrams.
Then, one assigns to the removed line the sum of the vectors associated with all vertices
in one of the two sub-diagrams.
Any use of the two sub-diagrams yields the same answer because of the constraint ${\vec k_1}+\cdots+{\vec k_n}=0$.

After associating the wavenumbers with all lines, we can assign the corresponding
factors to the vertices and the lines.
As for the vertex with p lines attached, assign the factor $N_{a1 \cdots a_p}$ to it.
As for the lines, assign $P$, where the argument of $P$ is set to the length of
the wavenumber associated with each line.
By multiplying all these factors assigned to vertices and lines, we obtain the corresponding
leading term.
Any term constructed in this way is given by a product of two terms.
One is the constant term which is given by contracting all the pairs of indices of 
the expansion coefficients assigned to each vertex of the tree diagram.
The other term is the $(n-1)$ product of the power spectra of the scalar field plus its permutations.
If we replace each power spectrum of the scalar fields by that of the curvature perturbation,
i.e. $P \to P_\zeta/(N_a N_a)$, the constant part gets a factor ${(N_a N_a)}^{n-1}$ in the denominator.
Following the cases for the bispectrum and the trispectrum,
we define the constant parameter corresponding to this diagram as a coefficient in front
of the $(n-1)$ product of $P_\zeta$.
By taking the sum over all the possible (connected) tree diagrams with $n$ vertices,
we obtain the $n$-point function.

It was shown in Ref.~\cite{Yokoyama:2008by} that the functions constructed from two tree
diagrams that are not isomorphic each other always yield different wavenumber dependence.
Therefore, the number of independent constant parameters required to specify the $n$-point 
function is equal to the one of all the possible connected tree diagrams with $n$ vertices that
are not isomorphic to each other.

A formula for a number $t_n$ of the independent connected tree diagrams with $n$ vertices is
given in Refs. \cite{Riordan:1958,Fry:1983cj}.
According to Refs. \cite{Riordan:1958,Fry:1983cj}, we can formally construct a function $t(x)$ as an infinite series,
\begin{equation}
t(x) = \sum_n t_n x^n.
\end{equation}
Then $t(x)$ is given by 
\begin{equation}
t(x)=r(x)-\frac{1}{2} r^2(x)+\frac{1}{2} r(x^2), \label{eqfort}
\end{equation}
where $r(x)$ is a function that satisfies,
\begin{equation}
r(x)=x \exp \bigg[ \sum_{k=1} \frac{1}{k} r(x^k) \bigg].
\end{equation}
From the last equation, we can recursively obtain the Taylor-expansion coefficients of $r(x)$.
Then from Eq.~(\ref{eqfort}), we can recursively obtain $t_n$ as well.
For example, $t_2=1,~t_3=1,~t_4=2,~t_5=3,~t_6=6,~t_7=11,~t_8=23,\cdots$.

\section{Inequalities of the Cauchy-Schwarz type}
\label{sec:inequality}

As we mentioned in the last section, wavenumber dependence of the $n$-point function is 
completely specified by a set of constant parameters each of which has a corresponding
connected tree diagram with $n$ vertices.
For the purpose of constructing the general inequalities among the non-linearity parameters,
it is convenient to adopt a normalization condition,
\begin{equation}
N_a N_a=1.
\end{equation}
Without a loss of generality, this condition can be always imposed by rescaling the
scalar field fluctuation.
With this condition, the denominator of the non-linearity parameters for the $n$-point 
function, which is given by ${(N_a N_a)}^{n-1}$, becomes unity.

\subsection{Definition of the non-linearity parameters}

For any connected tree diagram with $n$ vertices, if we cut a line appearing in the diagram, 
the diagram splits into two sub-diagrams.
If the resulting two sub-diagrams are isomorphic to each other, we call the parent
diagram symmetric diagram.
Obviously, in order for a diagram to be a symmetric diagram, $n$ must be even.
Let us denote by $\kappa_{2m}$ a number of all the symmetric diagrams among $t_{2m}$
tree diagrams with $2m$ vertices.
For example, $\kappa_2=1,~\kappa_4=1,~\kappa_6=2,~\kappa_8=4,~\kappa_{10}=9,\cdots$.
Our aim is to construct the inequalities of the Cauchy-Schwarz type for the non-linearity 
parameters up to the $n$-point functions.
As it will be clear later, this is possible when $n$ is even.
For the case of odd $n$, we can not construct the inequalities.
For example, there are inequalities among the non-linearity parameters up to the 
six-point function.
But there are no inequalities in which the highest order of the non-linearity 
parameters is five.
 
\begin{figure}[tbp]
  \begin{center}
    \includegraphics{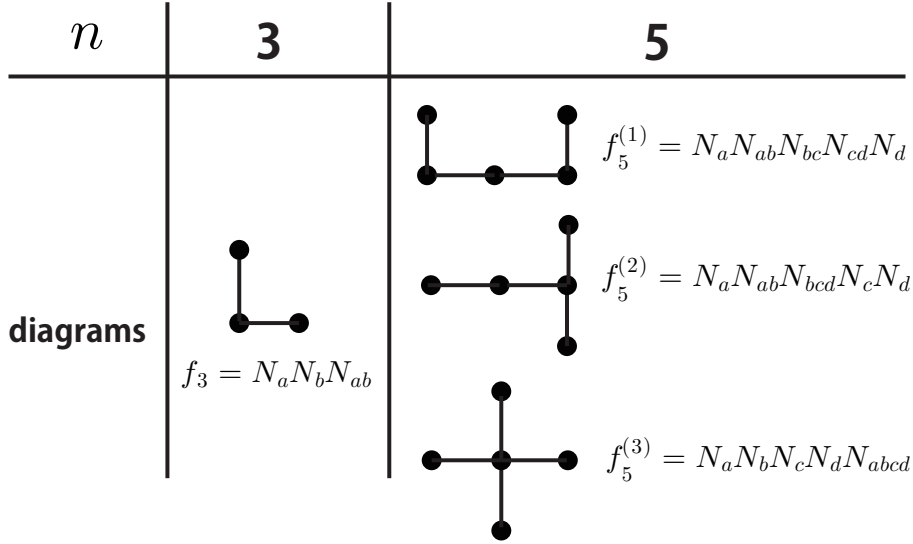}
  \end{center}
  \caption{The diagrams for $n=3,5$ cases.}
  \label{fig:three_five.eps}
\end{figure}

Since whether $n$ is even or odd is important, let us first write the non-linearity
parameters for the $2m$-point function as
\begin{equation}
\{ F_{2m}^{(1)},~\cdots,~F_{2m}^{(t_{2m})}\} =\{ \{ \tau_{2m}^{(1)},~\cdots,~\tau_{2m}^{(\kappa_{2m})} \},~\{g_{2m}^{(1)},~\cdots,~g_{2m}^{(t_{2m}-\kappa_{2m})} \} \}.
\end{equation}
Here, $\tau_{2m}^{(i)}$ and $g_{2m}^{(i)}$ are the non-linearity parameters 
for the symmetric diagram and the non-symmetric diagram, respectively.
\begin{figure}[tbp]
  \begin{center}
    \includegraphics{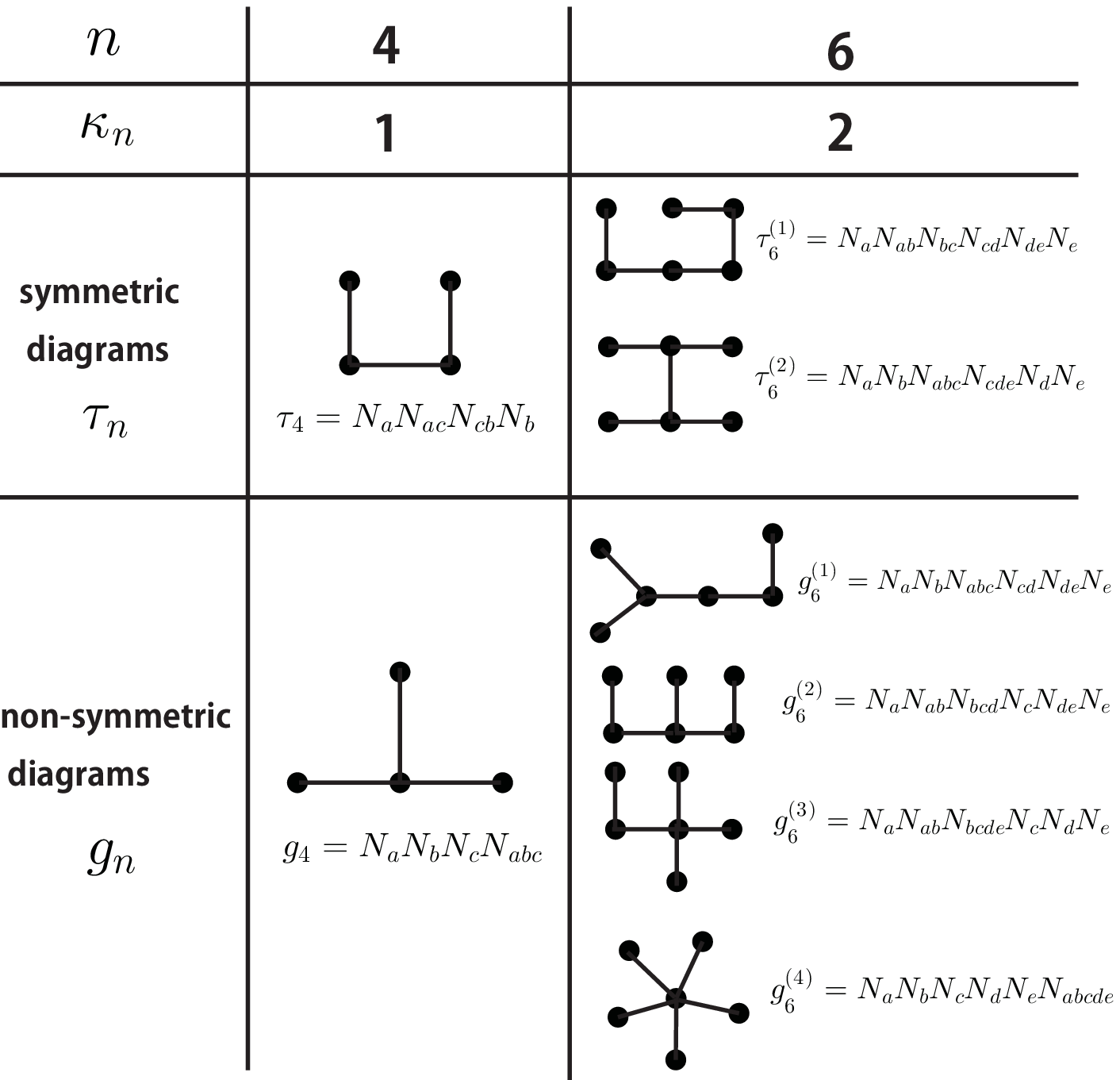}
  \end{center}
  \caption{The symmetric and non-symmetric diagrams for $n=4, 6$ cases.}
  \label{fig:symmetric_diagram.eps}
\end{figure}
In Fig.~\ref{fig:symmetric_diagram.eps},
we show the symmetric diagrams and the non-symmetric diagrams for $n=4,6$-point functions.
For the four-point function, a diagram corresponding to $\tau_{\rm NL}$ is a symmetric 
diagram while the one corresponding to $g_{\rm NL}$ is a non-symmetric diagram.
Therefore, relations between $(\tau_{4}^{(1)},~g_{4}^{(1)})$ and the conventional
non-linearity parameters $\tau_{\rm NL},~g_{\rm NL}$ are given by
\begin{equation}
\tau_{4}^{(1)}=\tau_{\rm NL},~~~~~g_{4}^{(1)}=\frac{54}{25} g_{\rm NL}.
\end{equation}
For $(2m-1)$-point function, we write the non-linearity parameters as
\begin{equation}
\{ F_{2m-1}^{(1)},~\cdots,~F_{2m}^{(t_{2m-1})}\} =\{ f_{2m-1}^{(1)},~\cdots,~f_{2m-1}^{(t_{2m-1})} \}.
\end{equation}
In Fig.~\ref{fig:three_five.eps},
we show the diagrams for $n=3,5$-point functions.
A relation between $f_{3}^{(1)}$ and the conventional non-linearity parameter
$f_{\rm NL}$ is given by
\begin{equation}
f_{3}^{(1)}=\frac{6}{5} f_{\rm NL}.
\end{equation}

\subsection{Construction of the inequalities}
Let us suppose that the highest order of the non-linearity parameters appearing
in the inequalities is $n=2m$.
We can construct the series of inequalities in a following way.

\subsubsection{The highest order $n=2m$}
Since $\tau_{2m}^{(i)}$ is a non-linearity parameter corresponding to the
symmetric diagram, it can be written as
\begin{equation}
\tau_{2m}^{(i)}={\vec V}_m^{(i)} \cdot {\vec V}_m^{(i)}, \label{highest1}
\end{equation}
where ${\vec V}_m^{(i)}$ is a vector associated with a diagram $D_m^{(i)}$
which is one of the sub-diagrams generated by dividing the symmetric diagram 
into two diagrams that are isomorphic to each other.
Therefore, $D_m^{(i)}$ is a tree diagram with $m$ vertices and with one external line.
Next, let us consider a quantity
\begin{equation}
{\vec V}_m^{(i)} \cdot {\vec V}_m^{(j)},
\end{equation}
for $i \neq j$. 
This must be a non-linearity parameter whose corresponding diagram is a diagram
with $2m$ vertices and is given by connecting the external line of $D_m^{(i)}$ 
with that of $D_m^{(j)}$.
Since this diagram is a non-symmetric diagram, there exists some $k$ such that
\begin{equation}
{\vec V}_m^{(i)} \cdot {\vec V}_m^{(j)}=g_{2m}^{(k)}. \label{highest2}
\end{equation}
Applying the Cauchy-Schwarz inequality to Eqs.~(\ref{highest1}) and (\ref{highest2}) 
yields an inequality,
\begin{equation}
\tau_{2m}^{(i)} \tau_{2m}^{(j)} \ge {\left( g_{2m}^{(k)} \right)}^2.
\end{equation}
A number of the independent inequalities of this kind is 
$ \kappa_{2m} \left( \kappa_{2m}-1 \right) / 2$.
These are the inequalities among the non-linearity parameters for $2m$-point function.

\subsubsection{Next highest order $2m-1$}
Let us then lower the order by one and consider a quantity,
\begin{equation}
{\vec V}_m^{(i)} \cdot {\vec V}_{m-1}^{(j)}.
\end{equation}
This is a non-linearity parameter whose corresponding diagram is obtained by
connecting the external line of $D_m^{(i)}$ with that of $D_{m-1}^{(j)}$. 
Therefore, there exists some $k$ such that
\begin{equation}
{\vec V}_m^{(i)} \cdot {\vec V}_{m-1}^{(j)}=f_{2m-1}^{(k)}. \label{nhighest1}
\end{equation}
Meanwhile, since a quantity ${\vec V}_{m-1}^{(j)} \cdot {\vec V}_{m-1}^{(j)}$
is a non-linearity parameter whose corresponding diagram is obtained by
connecting the external lines of the same two $D_{m-1}^{(j)}$,
we have 
\begin{equation}
{\vec V}_{m-1}^{(j)} \cdot {\vec V}_{m-1}^{(j)}=\tau_{2(m-1)}^{(j)}. \label{nhighest2}
\end{equation}
Applying again the Cauchy-Schwarz inequality to Eqs.~(\ref{nhighest1})
and (\ref{nhighest2}) yields
\begin{equation}
\tau_{2m}^{(i)} \tau_{2(m-1)}^{(j)} \ge {\left( f_{2m-1}^{(k)} \right)}^2.
\end{equation}
A number of the independent inequalities of this kind is 
$\kappa_{2m} \times \kappa_{2m-2}$.
These are the inequalities among the non-linearity parameters for $2m,~2m-1$ and
$2m-2$-point functions.

\subsubsection{Next-to-next highest order $2m-2$}
Let us further lower the order by one and consider a quantity
\begin{equation}
{\vec V}_m^{(i)} \cdot {\vec V}_{m-2}^{(j)}.
\end{equation}
This is a non-linearity parameter whose corresponding diagram is obtained by
connecting the external line of $D_m^{(i)}$ with that of $D_{m-2}^{(j)}$. 
Since this diagram can be both symmetric or non-symmetric, we write it as
\begin{equation}
{\vec V}_m^{(i)} \cdot {\vec V}_{m-2}^{(j)}=F_{2m-2}^{(k)}.
\end{equation}
Applying the same reasoning as before, we obtain
\begin{equation}
\tau_{2m}^{(i)} \tau_{2(m-2)}^{(j)} \ge {\left( F_{2m-2}^{(k)} \right)}^2.
\end{equation}
A number of the independent inequalities of this kind is 
$\kappa_{2m} \times \kappa_{2m-4}$.
These are the inequalities among the non-linearity parameters for $2m,~2(m-1)$ and
$2(m-2)$-point functions.

By repeating the above procedures until the order reduces to one, 
we can obtain all the Cauchy-Schwarz type inequalities that involve the
non-linearity parameters for the $2m$-point function as the highest order.
In particular, at the final procedure, we end up with getting a set of inequalities:
\begin{equation}
\tau_{2m}^{(i)} \ge {\left( f_{m+1}^{(k)} \right)}^2.
\end{equation}
A number of the independent inequalities of this kind is $\kappa_{2m}$.

Combining all these results, a number $a_{2m}$ of independent inequalities 
that involve the non-linearity parameters for the $2m$-point function as the highest order
is given by
\begin{equation}
a_{2m}=\kappa_{2m} \left( \frac{1}{2}(\kappa_{2m}-1)+\kappa_{2m-2}+\kappa_{2m-4}+\cdots + \kappa_2 \right). 
\end{equation}
For example, $a_4=1,~a_6=5,~a_8=22,~a_{10}=108,\cdots$.

\subsection{Application to four, six and eight-point functions}
\subsubsection{Case of four-point function}

There are two different tree diagrams for the four-point function.
One of them is a symmetric diagram which corresponds to $\tau_{\rm NL}$ and 
the other one is a non-symmetric diagram corresponding to $g_{\rm NL}$.
According to the general argument we developed in the last subsection,
there is no inequality between $\tau_{\rm NL}$ and $g_{\rm NL}$ because $\kappa_4=1$.
Then, let us lower the order by one. At this stage, we obtain one inequality
\begin{equation}
\tau_4 \ge f_3^2.
\end{equation}
In terms of the conventional non-linearity parameters, this can be written as
\begin{equation}
\tau_{\rm NL} \ge \frac{36}{25} f_{\rm NL}^2.
\label{eq:ineq_4}
\end{equation}
Since we have reduced to the lowest order, there are no other inequalities.
%
%
%
%

\subsubsection{Case of six-point function}
Since $\kappa_6=2$, there exists a single inequality among the non-linearity
parameters for the six-point function:
\begin{equation}
\tau_6^{(1)} \tau_6^{(2)} \ge {\left( g_6^{(1)} \right)}^2.
\label{eq:ineq_61}
\end{equation}
where the correspondence between the diagrams and the non-linearity parameters is
given in Fig.~\ref{fig:symmetric_diagram.eps}.
By lowering the order by one, we obtain $\kappa_6 \times \kappa_4 =2$ inequalities:
\begin{equation}
\tau_6^{(1)} \tau_4 \ge {\left( f_5^{(1)} \right)}^2,~~~~~\tau_6^{(2)} \tau_4 \ge {\left( f_5^{(2)} \right)}^2.
\label{eq:ineq_62}
\end{equation}
Reducing further the order by one, we obtain $\kappa_6 \times \kappa_2=2$ inequalities:
\begin{equation}
\tau_6^{(1)} \ge \tau_4^2,~~~~~\tau_6^{(2)} \ge g_4^2.
\label{eq:ineq_63}
\end{equation}
The last two inequalities were also provided in \cite{Lin:2010ua}.
Since we have reduced to the lowest order, there are no other inequalities.
Total number of inequalities is five.
\subsubsection{Case of eight-point function}
\begin{figure}[htbp]
  \begin{center}
    \includegraphics[keepaspectratio=true,height=150mm]{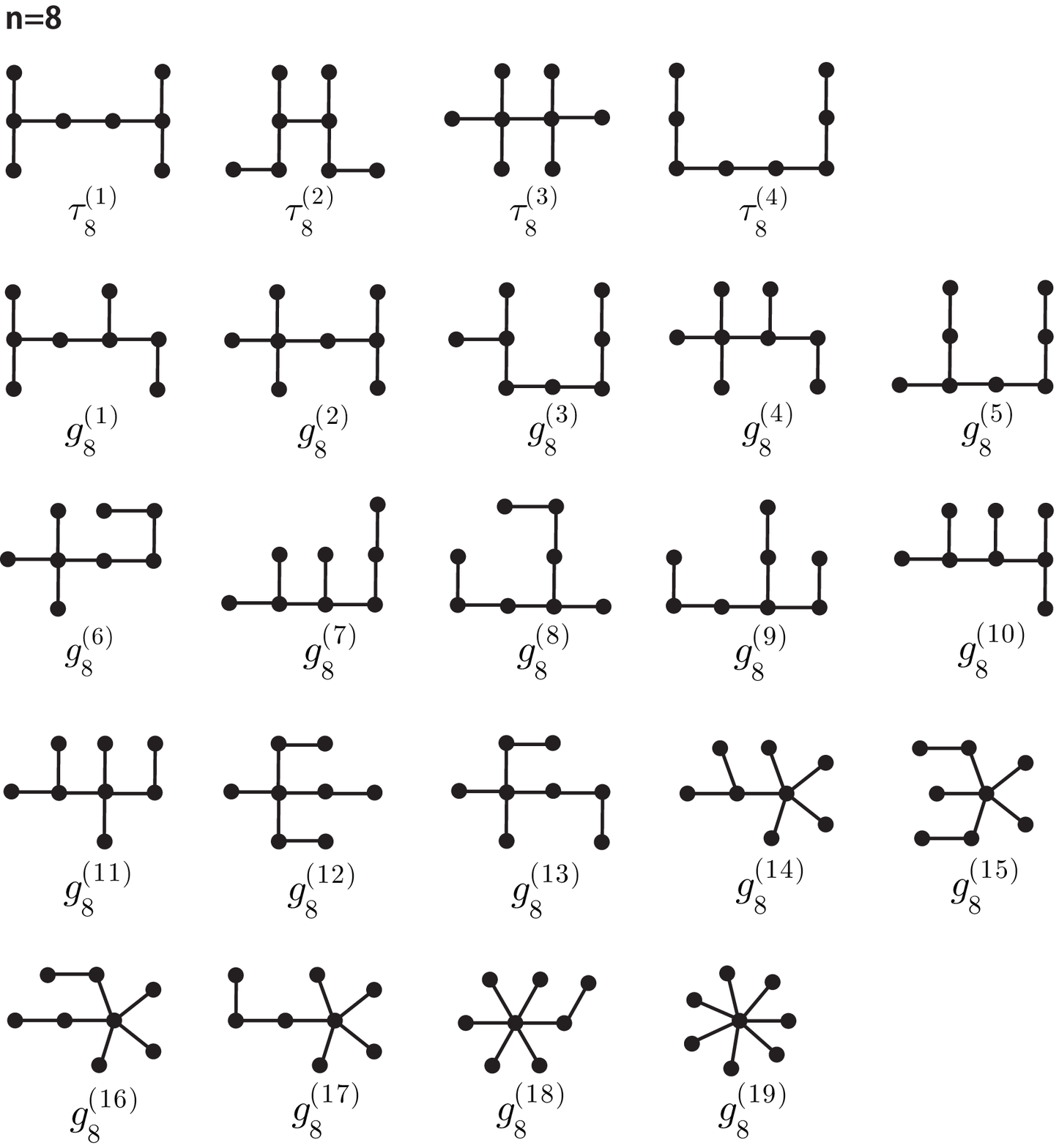}
  \end{center}
  \caption{The correspondence between the diagrams and the non-linearity parameters for $8$-point function.}
  \label{fig:n8.eps}
\end{figure}

\begin{figure}[htbp]
  \begin{center}
    \includegraphics{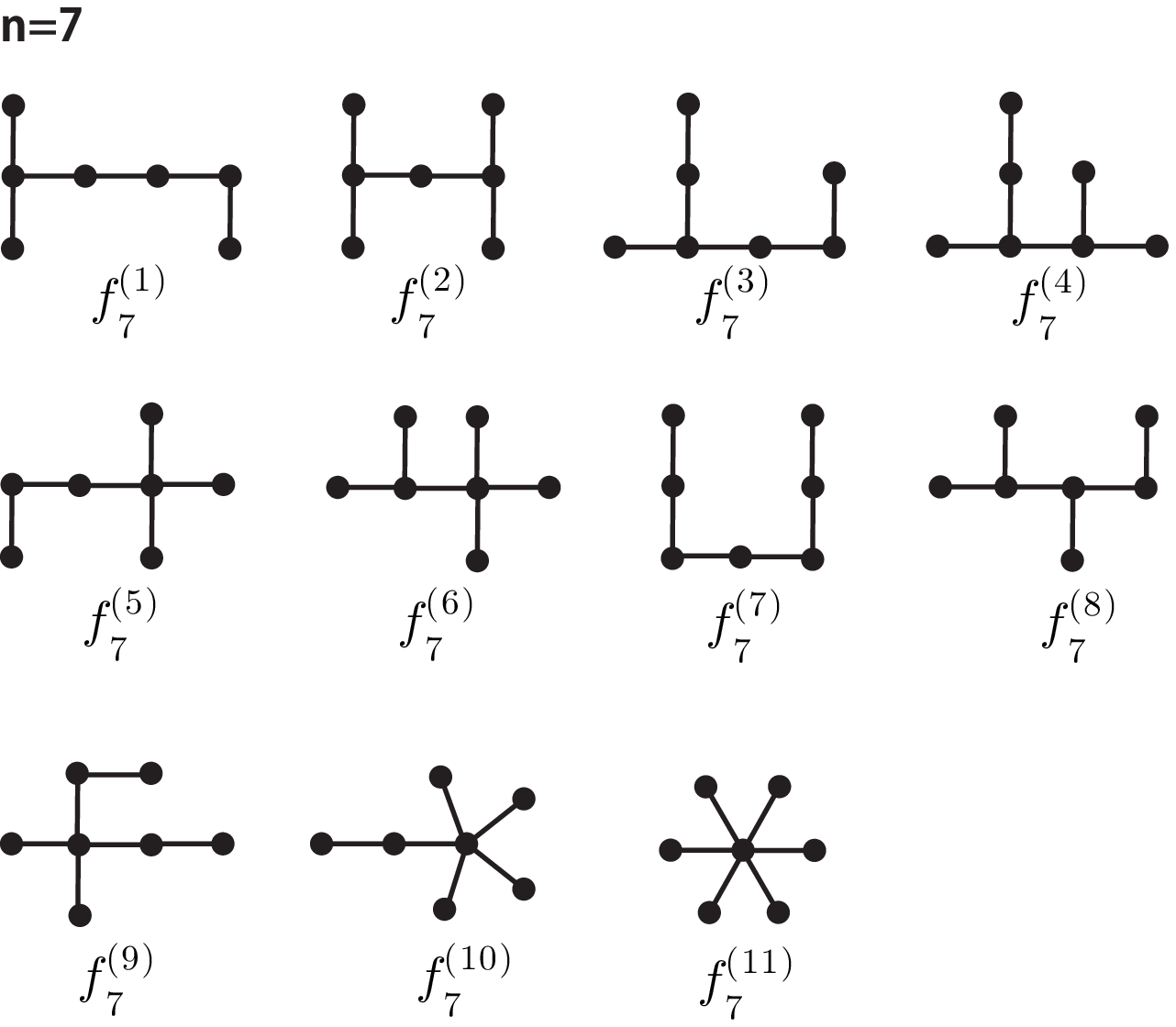}
  \end{center}
  \caption{The correspondence between the diagrams and the non-linearity parameters for $7$-point function.}
  \label{fig:n7.eps}
\end{figure}

Since $\kappa_8=4$, there exist six inequalities among the non-linearity
parameters for the eight-point function:
\begin{eqnarray}
&& \tau_8^{(1)} \tau_8^{(2)} \ge {\left( g_8^{(1)} \right)}^2,~~~\tau_8^{(1)} \tau_8^{(3)} \ge {\left( g_8^{(2)} \right)}^2,~~~\tau_8^{(1)} \tau_8^{(4)} \ge {\left( g_8^{(3)} \right)}^2, \nonumber \\
&& \tau_8^{(2)} \tau_8^{(3)} \ge {\left( g_8^{(4)} \right)}^2,~~~\tau_8^{(2)} \tau_8^{(4)} \ge {\left( g_8^{(5)} \right)}^2,~~~\tau_8^{(3)} \tau_8^{(4)} \ge {\left( g_8^{(6)} \right)}^2.
\end{eqnarray}
where the correspondence between the diagrams and the non-linearity parameters is
given in Fig.~\ref{fig:n8.eps}.
By lowering the order by one, we obtain $\kappa_8 \times \kappa_6 =8$ inequalities:
\begin{eqnarray}
\tau_8^{(1)} \tau_6^{(1)} \ge {\left( f_7^{(1)} \right)}^2,~~~\tau_8^{(1)} \tau_6^{(2)} \ge {\left( f_7^{(2)} \right)}^2,~~~\tau_8^{(2)} \tau_6^{(1)} \ge {\left( f_7^{(3)} \right)}^2,~~~\tau_8^{(2)} \tau_6^{(2)} \ge {\left( f_7^{(4)} \right)}^2, \nonumber \\
\tau_8^{(3)} \tau_6^{(1)} \ge {\left( f_7^{(5)} \right)}^2,~~~\tau_8^{(3)} \tau_6^{(2)} \ge {\left( f_7^{(6)} \right)}^2,~~~\tau_8^{(4)} \tau_6^{(1)} \ge {\left( f_7^{(7)} \right)}^2,~~~\tau_8^{(4)} \tau_6^{(2)} \ge {\left( f_7^{(1)} \right)}^2 \nonumber, 
\end{eqnarray}
where the correspondence between the diagrams and the non-linearity parameters is
given in Fig.~\ref{fig:n7.eps}.
Reducing further the order by one, we obtain $\kappa_8 \times \kappa_4=4$ inequalities:
\begin{eqnarray}
\tau_8^{(1)} \tau_4 \ge {\left( g_6^{(1)} \right)}^2,~~~\tau_8^{(2)} \tau_4 \ge {\left( g_6^{(2)} \right)}^2,~~~\tau_8^{(3)} \tau_4 \ge {\left( g_6^{(3)} \right)}^2,~~~\tau_8^{(4)} \tau_4 \ge {\left( \tau_6^{(1)} \right)}^2.
\end{eqnarray}
Reducing further the order by one, we obtain $\kappa_8 \times \kappa_2=4$ inequalities:
\begin{eqnarray}
\tau_8^{(1)} \ge {\left( f_5^{(2)} \right)}^2,~~~\tau_8^{(2)} \ge {\left( f_5^{(2)} \right)}^2,~~~\tau_8^{(3)} \ge {\left( f_5^{(3)} \right)}^2,~~~\tau_8^{(4)} \ge {\left( f_5^{(1)} \right)}^2.
\end{eqnarray}
Since we have reduced to the lowest order, there are no other inequalities.
Total number of inequalities is 22.

\section{Conclusion}
\label{sec:con}

The primordial non-Gaussianity has been focused on by many authors as a new probe of the inflation
dynamics.
In addition to the bispectrum, use of the higher order correlation functions will become useful 
in the future. 
In particular, the inequality between $\tau_{\rm NL}$ and $g_{\rm NL}$ shows the importance of the trispectrum
to look for the non-Gaussianity in the primordial perturbations.

In this paper, we developed a general formalism to construct the Cauchy-Schwarz type inequalities  
among the non-linearity parameters for the higher order correlation functions.
This method enables us to derive all the inequalities up to any-point function.
As an application, we explicitly derived all the inequalities up to the four, six and eight-point
functions.
We first confirmed that there is just one inequality, which is Eq.~(\ref{ft}), up to the four-point function.
Up to the six-point function, there are five new inequalities two of which were given in Ref. \cite{Lin:2010ua}.
Up to the eight-point function, there are further 22 new inequalities.\\

\noindent {\bf Acknowledgments:} 
T.~S. is supported by a Grant-in-Aid for JSPS Fellows No.~1008477.
S.~Y. is partially supported by the
Grant-in-Aid for Scientific research from the Ministry of Education,
Science, Sports, and Culture, Japan, No. 22340056.
S.~Y. also acknowledges support from the Grant-in-Aid for
the Global COE Program ``Quest for Fundamental Principles in the
Universe: from Particles to the Solar System and the Cosmos'' from
MEXT, Japan. 
The authors also thank the Yukawa Institute for Theoretical Physics at Kyoto University,
where this work was discussed during the YITP-T-10-05 on "Cosmological Perturbation and
Cosmic Microwave Background".

\appendix


\end{document}